\documentstyle[prb,aps,epsfig,floats,amsfonts]{revtex}

\begin{document}

\twocolumn \psfull \draft

\wideabs{
\title{Reduction of Josephson critical current in short ballistic $SNS$ weak links}
\author{Branislav K. Nikoli\' c,$^a$ J. K. Freericks,$^a$ and P. Miller$^b$}
\address{$^a$Department of Physics, Georgetown University,
Washington, DC 20057-0995 \\
$^b$Department of Physics, Brandeis University, Waltham, MA 02454}

\maketitle

\begin{abstract}
We present fully self-consistent calculations of the thermodynamic
properties of three-dimensional clean $SNS$ Josephson junctions,
where $S$ is an $s$-wave short-coherence-length superconductor and
$N$ is a clean normal metal. The junction is modeled on  an
infinite cubic lattice such that the transverse width of the $S$ is
the same as that of the $N$, and its thickness is tuned from the short
to long limit. Both the reduced order parameter near the $SN$ boundary 
and the short coherence length depress the critical Josephson 
current $I_c$, even in short junctions. This is contrasted with recent 
measurements on $SNS$ junctions finding much smaller $I_cR_N$ products than 
expected from the standard (non-self consistent and quasiclassical) 
predictions. We also find unusual current-phase relations, a 
``phase anti-dipole'' spatial distribution of the self-consistently determined
contribution to the macroscopic phase, and an ``unexpected'' minigap in the local 
density of states within the $N$ region. 
\end{abstract}

}

\narrowtext

Over the past decade, both experimental and theoretical interest in 
the  superconductivity of inhomogeneous systems have been rekindled, 
thereby leading to a reexamination of even well-charted areas from 
the mesoscopic point of view.~\cite{superlatt} For example, 
the Josephson effect in a superconductor--normal-metal--superconductor ($SNS$) 
weak link was known to be the result of the macroscopic condensate wave function 
leaking from the $S$ into the $N$ region. The induction of such superconducting 
correlations in the $N$, the so-called proximity effect, has been given a new 
real-space interpretation through the relative phase-coherence of quasiparticles, 
correlated by Andreev reflection at the $SN$ interface.~\cite{andreev} 
Moreover, the realization of the importance of tracking the phase-coherence of 
single particle wave-functions in proximity-coupled metals of mesoscopic size 
has also unearthed new phenomena, such as quantization of the critical current 
in ballistic mesoscopic short $SNS$ junctions at low enough temperature.~\cite{carlo_spc,furusaki} 
In short clean junctions, as $T \rightarrow 0$, the critical supercurrent 
$I_c = e\Delta/\hbar$ carried by a single conducting channel depends only on  
the superconducting energy gap $\Delta$ as the smallest energy scale $\Delta < E_{\rm Th}=\hbar v_F^N/L$ 
(in the long junction limit $I_c \sim E_{\rm Th}$ is set~\cite{long} by the 
``ballistic'' Thouless energy $E_{\rm Th} < \Delta$, which is a single 
quasiparticle property determined by the Fermi velocity $v_F^N$ in the $N$ 
interlayer of length $L$). Thus, both mesoscopic and  ``classical''  
clean point-contact $SNS$ junctions, with ballistic transport $\ell > L$ ($\ell$ is 
the mean free path), are predicted to exhibit the same $I_cR_N=\pi \Delta/e$ product 
at $T=0$. This has been known for quite some time as the Kulik-Omelyanchuk (KO) 
formula,~\cite{ko77}  where $R_N$ is the Sharvin point contact resistance 
$R_N=h/2e^2M$  of the ballistic $N$ region containing $M$ conducting channels. 

Recent experimental activity~\cite{klapwijk} on highly transparent~\cite{footnote} ballistic 
short $SNS$ junctions, where both $I_c$ and $R_N$ are independent of the 
junction length, reveals much lower values of $I_cR_N$  than the KO formula 
(similarly, the critical current steps found in an attempt~\cite{taka} to observe 
discretized $I_c$ are much smaller than the predicted $e\Delta/\hbar$). However, a proper 
interpretation of these results demands a clear understanding of the relationship 
between relevant energy and length scales. The criterion for the short 
junction limit $\Delta < E_{\rm Th}$  introduces a  ``coherence length'' of the junction  
$\xi_0=\hbar v_F^N/\pi \Delta$; i.e., the maximum KO limit can be expected only for 
$L \ll \xi_0$.  The relation between $k_B T$ and $E_{\rm Th}$ defines the 
high- ($k_B T > E_{\rm Th}$) versus low- ($k_B T < E_{\rm Th}$) temperature limits, 
which is equivalently expressed in terms of the junction thickness as $L > \xi_N$ 
versus $L < \xi_N$, respectively,  with $\xi_N=\hbar v_F^N/2\pi k_B T$ being the normal 
metal coherence length. The $\xi_N$ sets the scale over which two quasiparticles in 
the $N$, correlated  by Andreev reflection, retain their relative phase coherence 
(i.e., superconducting correlations imparted on the $N$ region at finite temperature 
decay exponentially with $\xi_N$, while at zero temperature $\xi_N \rightarrow \infty$ 
and the condensate wave function decays inversely in the distance from the interface~\cite{falk}). 
Therefore, the simple exponential decay of $I_c \sim \exp(-L/\xi_N)$ appears only in 
the high-temperature limit, while in the opposite low-temperature limit $\xi_N$ ceases 
to be a relevant length scale and the decay is slower than exponential. The aforementioned 
experiments on clean $SNS$ junctions~\cite{klapwijk} are conducted on Nb/InAs/Nb junctions 
which are tuned to lie in the regime where $\xi_S < L < \xi_0 \ll \xi_N$ ($\xi_S=\hbar v_F^S/\pi 
\Delta$ is the bulk superconductor coherence length). Thus, the large difference between $\xi_S$ 
and $\xi_0$ means that there is a substantial Fermi velocity mismatch (typically an order of 
magnitude~\cite{klapwijk,taka}), which must generate normal scattering at the $SN$ interface
in addition to Andreev reflection. This, together with other possible sources of scattering at 
the $SN$ boundary, like imperfect interfaces~\cite{furusaki,iscatter,chrestin} or charge accumulation 
layers~\cite{sinis} (typical of Nb/InAs contact), cannot be detected by only observing the 
independence of $I_c$ and $R_N$ on interelectrode separation (for intermediate~\cite{klapwijk} $L$). 
Nevertheless, this is frequently the criterion used in experiments~\cite{klapwijk} to ensure that 
the transport is ballistic. Therefore, the ideal maximum value for $I_c$ could be achieved only for 
$\xi_S=\xi_0$ ($v_F^S=v_F^N$) and with a perfectly transparent interface, where 
the junction thickness satisfies $L \ll \xi_S$. Even in this case it is possible that the current 
in short junctions is smaller than expected due to a depressed value 
of the order parameter on the $SN$ boundary when the transverse width of the $S$ and 
$N$ regions are the same.~\cite{sols} Such junctions cannot be treated by simplified 
approaches~\cite{carlo_spc,furusaki} assuming a step function for $\Delta(z)$ because 
the order parameter varies within the $S$ due to the self-consistency.~\cite{sols,levy}
\begin{figure}
\centerline{\psfig{file=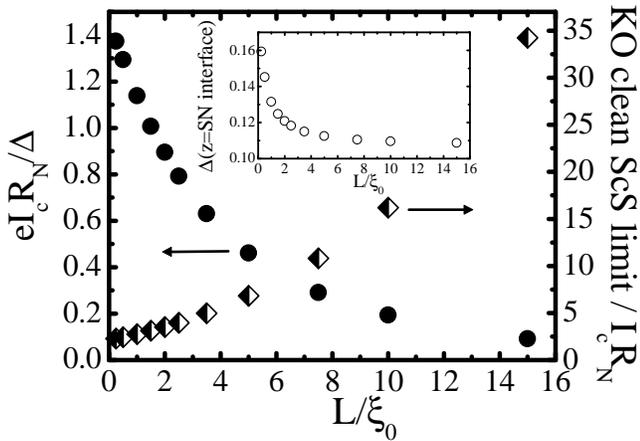,height=3.3in,angle=-90} }
\vspace{0.2in} \caption{Product of the critical current $I_c$ and
the normal state resistance $R_N$ as a function of the $SNS$
junction thickness $L$. Both the $S$ and $N$ are at half-filling in
the bulk. The value of $I_cR_N$ is always below the product of the bulk
critical current in the $S$ leads and the Sharvin point contact
resistance, $I_c^{\rm bulk} R_{\rm Sh}=1.45 \Delta/e$.
The right axis measures the ratio of the Kulik-Omelyanchuk 
formula $I_cR_N=\pi\Delta/e$ for the clean  superconducting point 
contact ($L \rightarrow 0$) and $I_cR_N$ of our junctions. The inset shows 
the decay of the order parameter at the $SN$ interface for $I=0$, 
which reaches an asymptotic value of about one-half of the bulk $\Delta$ 
for $L \simeq 2 \xi_S$ [$I_cR_N/\Delta(z=SN \ {\rm interface})$ 
is virtually constant for $L < 2 \xi_S$].} 
\label{fig:icrn_sns}
\end{figure}

Here we undertake an idealized study of different {\em intrinsic}
properties of three-dimensional $SNS$ junctions which can be detrimental 
to $I_c$, without invoking any sample-fabrication dependent additional 
scattering at the $SN$ interface. Two such effects are known: (i) the requirements 
of self-consistency, which become important for specific junction geometries 
delineated below, depresses the order parameter  near the $SN$ boundary and 
therefore the current in short junctions; (ii) a finite ratio
$\Delta/\mu$ (where $\mu$ is the Fermi energy measured from the 
bottom of the band) generates intrinsic  normal scattering at the 
$SN$ boundary (without the presence of impurities or barriers 
at the interface). 
\begin{figure}
\centerline{\psfig{file=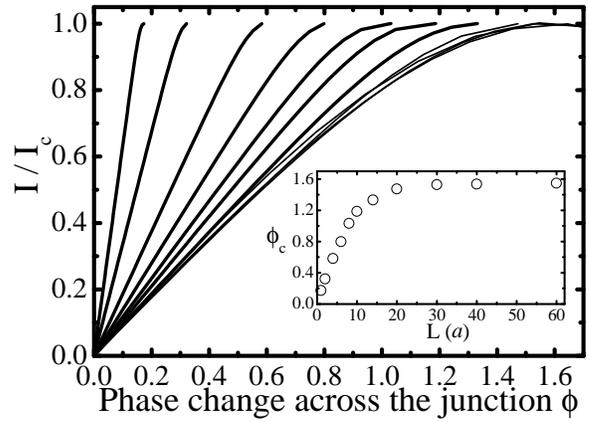,height=3.0in,angle=-90} }
\vspace{0.2in} \caption{Scaling of the current-phase relation
$I(\phi)/I_c$ with the thickness of a clean $SNS$ junction. Note
that the phase change across the junction $\phi_c$ at the 
critical current $I_c=I(\phi_c)$ varies monotonically with the 
junction thickness, as shown in the inset, and is always far 
below $\pi$, which is the prediction of non-self-consistent calculations 
in both the short~\cite{ko77} ($L \ll \xi_0$)  and 
long~\cite{long} ($L \gg \xi_0$) junction limits at $T \rightarrow 0$.} 
\label{fig:cvp_sns}
\end{figure}
Therefore, even a clean junction (with ballistic 
transport above $T_c$) might not be in the ballistic limit~\cite{hurd} 
below the superconducting transition temperature $T_c$, unless the filling 
is tuned to the energy of the transmission resonances. Our principal result 
for the evolution of $I_cR_N$ as a function of $L$ is shown in 
Fig.~\ref{fig:icrn_sns}. The $I_cR_N$ drops by about an order of 
magnitude at $L \sim \xi_N$, thus showing how the characteristic 
voltage can be reduced dramatically in moderate length junctions, even in the 
low-temperature limit (to which our junctions belong). The reduction of $I_c$ in our 
short junctions is determined by the depression of the order parameter in the $S$, 
as demonstrated by the inset in Fig.~\ref{fig:icrn_sns} where $\Delta(z)$ at the $SN$ 
interface decreases asymptotically to a limiting value reached at $L \gtrsim 2\xi_S$ 
with $I_cR_N/\Delta(z=SN \ {\rm interface})$ being nearly a constant for $L < 2 \xi_S$. 
For the junctions thicker than $2\xi_S$ the decay of the critical current $I_c \sim 1/L$ scales 
as~\cite{long} $E_{\rm Th}$, while at non-zero temperatures and for long enough junctions 
$L > \xi_N$ it changes into a simple exponential decay. Thus, in the general 
case $\xi_S < \xi_N$, $I_c$ can be independent of $L$ only for $2 \xi_S < L <\xi_0$, as observed 
in the experiments. However, 
such thickness-independent $I_c$ can be substantially below $M e\Delta /\hbar$, as defined 
by the inevitable ($v_F^S \neq v_F^N$) interface scattering and/or reduced  $\Delta$, with 
its lowest value being set at $L \simeq 2\xi_S$ by the ``inverse proximity effect'' on the $S$ 
side of a $SN$ structure. We believe that ballistic behavior could be found in our junctions 
at even lower $T$, where $\xi_N \gg 2 \xi_S$, but such calculations are technically more 
involved at present.

The $SNS$ Josephson junction is modeled by a Hubbard Hamiltonian
\begin{eqnarray} \label{eq:hamiltonian}
H&=&-\sum_{ij\sigma}t_{ij}c_{i\sigma}^{\dag}c_{j\sigma} \nonumber \\
&& \mbox{}+\sum_iU_i\left (
c_{i\uparrow}^{\dag}c_{i\uparrow}-\frac{1}{2}\right ) \left (
c_{i\downarrow}^{\dag}c_{i\downarrow}-\frac{1}{2}\right ),
\end{eqnarray}
on a simple cubic lattice (with lattice
constant $a$). Here $c_{i\sigma}^{\dag}$ ($c_{i\sigma}$) creates
(destroys) an electron of spin $\sigma$ at site $i$, $t_{ij}=t$
(the energy unit) is the hopping integral between nearest
neighbor sites $i$ and $j$. The $SNS$ structure is comprised 
\begin{figure}
\centerline{\psfig{file=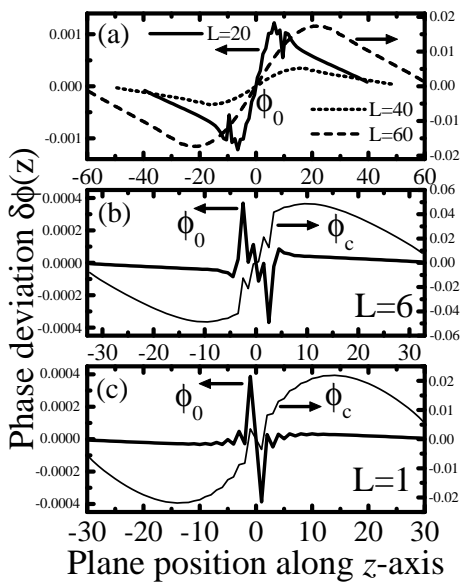,height=3.0in,angle=0} }
\vspace{0.2in} \caption{Scaling of a spatial distribution of the 
phase deviation $\delta \phi(z)$ within the self-consistently 
modeled part of the clean $SNS$ junction. The total phase change 
across the junction $\phi$ is the sum of the bulk phase gradient$\times L$ 
and the change in $\delta \phi(z)$ along the $N$ interlayer, 
Eq.~(\ref{eq:phasechange}): $\phi_0$ at some small supercurrent 
and $\phi_c$ at the critical junction current $I_c$. At large enough 
junction thickness $L$ [panel (a)] the shape of $\delta \phi(z)$ is just
rescaled by the increase of the Josephson current, while 
for smaller $L$ the shape changes abruptly 
upon  approaching $I_c$ [panels (b) and (c)].} \label{fig:phase_sns}
\end{figure}
of stacked planes~\cite{miller} where $U_{i}<0$ is the attractive 
interaction for sites within the superconducting planes (inside the $N$ 
region $U_i=0$). In the Hartree-Fock approximation (HFA), this leads to a BCS 
mean-field superconductivity in the $S$ leads, where for $U_i=-2$ and 
half-filling we get $\Delta =0.197t$  ($T_c=0.11t$) and 
$\xi_S=\hbar v_F^S/\pi \Delta \simeq 4a$. The lattice Hamiltonian~(\ref{eq:hamiltonian}) 
of the inhomogeneous $SNS$ system is solved by computing a Nambu-Gor'kov matrix 
Green function. The off-diagonal block of this matrix is the anomalous average which quantifies 
the establishment of superconducting correlations in either the $S$ 
[$\Delta(z) = |U(z)| F(z)$, where $F(z_i,z_i,\tau=0^+)$ is the pair-field amplitude] or the 
$N$ region. For the local interaction treated in the HFA, computation of the Green function 
reduces to inverting an infinite block tridiagonal Hamiltonian matrix in
real space. The Green functions are thereby expressed through a matrix
continued fraction (technical details are given elsewhere~\cite{miller}). 
The final solution is fully self-consistent in the order parameter 
$|\Delta(z)|e^{i \phi(z)}$ inside the part of the junction comprised of 
the $N$ region and the first 30 planes inside the superconducting 
leads on each side of the $N$ interlayer. 
\begin{figure}
\centerline{\psfig{file=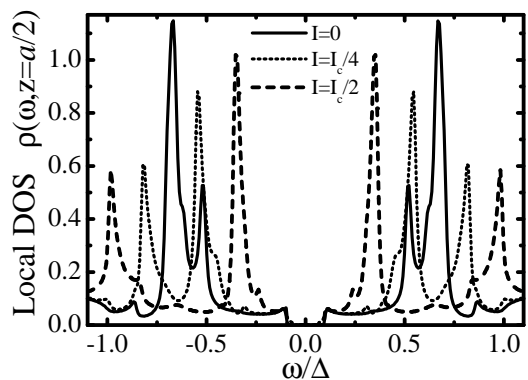,height=2.7in,angle=-90} }
\vspace{0.2in} \caption{Local density of states at the central
plane  of a clean 3D $SNS$ junction (composed of 10 normal
planes, $L=10a$) for different supercurrent flows: $I=0$, $I=I_c/4$, 
and $I=I_c/2$. The peaks correspond to ABS confined within the $N$ region 
at energies $E < \Delta$. For $I \neq 0$, the degeneracy of right- and left-moving 
electrons is lifted by a Doppler shift, giving rise to the 
Josephson DC current (or at least part of it~\cite{long,bagwell}). 
The minigap around the Fermi energy $\omega=0$ in the $N$ 
region appears to be a result of a finite $\Delta/\mu \approx 0.03$ generating 
normal scattering at the $SN$ interfaces.} 
\label{fig:ldos_sns}
\end{figure}
Our Hamiltonian formulation of
the problem and its solution by this Green function technique is
equivalent to solving a discretized version of the Bogoliubov-de
Gennes~\cite{dege} (BdG) equations formulated in terms of Green functions,~\cite{bagwell} 
but in a fully self-consistent  manner---by  determining the off-diagonal 
pairing potential $\Delta(z)$ in the BdG Hamiltonian~\cite{levy} after each, 
iteration until convergence is achieved. The tight-binding description of 
the electronic states also allows us to include an arbitrary band structure 
(which is rarely taken into account~\cite{rusi_band}), or more complicated 
pairing symmetries. The calculation is performed at $T=0.09T_c$ where $\xi_N=40a$, 
which is a low-temperature limit for almost all of our junction thicknesses.

This technique is different from the quasiclassical use of
a coarse-grained microscopic Gor'kov Green function, through
either the Eilenberger equations (clean limit) or Usadel equations
(dirty limit),~\cite{schon} or non-self-consistent solutions of the BdG
equations~\cite{carlo_spc} which are applicable only for special
geometries where the left and right $S$ leads can be characterized
by a constant phase $\phi_L$ and $\phi_R$, respectively. This
neglects the phase gradient $(d \phi/dz)_{\rm bulk}$ inside the
$S$, thereby violating current conservation. Such an assumption is justified 
when the critical current of the junction is limited by,
e.g., a point contact geometry, which requires a much smaller
gradient than $1/\xi_S$ at the critical current density in the
bulk, while the Josephson current is determined by the region within
$\xi_S$ from the junction.~\cite{carlo_spc}
Since we choose the $S$ and $N$ layers of the same transverse
width, $I_c/I_c^{\rm bulk}$ can be close to one for thin
junctions. In such cases, current flow affects appreciably the
superconducting order parameter [i.e., $F(z)$ both inside and
outside the $N$] and a self-consistent treatment becomes
necessary (as is the general case of finding the critical current
of a bulk superconductor~\cite{bagwell94,bardeen}). Since for a
clean $SNS$ junction $R_Na^2=[(2e^2/h)(k_F^2 /4\pi)]^{-1} \approx
1.58 ha^2/2e^2$ is just the Sharvin point contact resistance
(i.e., inverse of the conductance, at half-filling, per unit area
$a^2$ of our junction with infinite cross section), the absolute
limit of the characteristic voltage is  $I_c^{\rm bulk} R_{\rm
N}=1.45 \Delta/e$ set by the bulk critical current $I_c^{\rm
bulk}$ of the $S$ leads,~\cite{bagwell94} as shown in
Fig.~\ref{fig:icrn_sns}. In three-dimensional (3D) junctions
$I_c^{\rm bulk}= 1.09 e n \Delta/\hbar k_F $ (per unit area
$a^2$, at half-filling) is slightly higher than the current
density determined by the Landau depairing velocity $v_{\rm
depair}=\Delta/\hbar k_F$, at which superfluid flow breaks the phase
coherence of Cooper pairs,~\cite{bagwell94} because of the
possibility of gapless superconductivity at superfluid velocities
slightly exceeding~\cite{bardeen} $v_{\rm depair}$. Although our
$I_cR_N$ is always smaller than the ideal KO limit, it is still above 
the experimentally measured values~\cite{klapwijk} in the intermediate junction
thicknesses, which are about hundred times smaller
than the KO limit. This suggests that additional scattering
confined to the interface region is indeed necessary to account
for such small values.~\cite{iscatter,sinis}

Since self-consistent calculations require a phase gradient
inside the $S$ (which we choose to be a boundary condition in the
bulk of the superconductor), we must carefully define how to parameterize 
the Josephson current. There are two possibilities: either a global 
phase change across the $N$ region~\cite{kupriyanov92} or the phase 
offset~\cite{sols} which is related to the phase change by a nontrivial scale
transformation. We use a global phase change which in a discrete model 
like~(\ref{eq:hamiltonian}) requires a convention. The thickness
of the junction is defined to be the distance measured from the
point $z_L$, in the middle of the last $S$ plane on the left (at
$z_L^S$) and the first adjacent $N$ plane (at $z_L^N=z_L^S+1$),
to the middle point $z_R$ between the last $N$ and first $S$
plane on the right. Since $\phi(z)$ is defined within the planes,
we set $\phi(z_L)=[\phi(z_L^S)+\phi(z_L^N)]/2$ to be the phase at
$z_L$, and equivalently for $\phi(z_R)$. The phase change across
the barrier is then given by 
\begin{equation}\label{eq:phasechange}
\phi = L\left( \frac{d\phi}{dz} \right)_{\rm bulk} + 
\delta \phi(z_R)- \delta \phi(z_L), 
\end{equation} 
where $\delta \phi(z)$ is the 
``phase deviation'' which develops self-consistently on  top of the 
imposed linear background variation of the phase. The current versus phase 
change relation is plotted in Fig.~\ref{fig:cvp_sns}. Non-self 
consistent calculations predict that $I_c$ occurs at $\phi_c=\pi$ for 
both~\cite{ko77} $ScS$ and long $SNS$ junctions (at $T=0$).~\cite{long} 
However, the self-consistent analysis leads 
to a sharp deviation from these notions,~\cite{sols} which is most conspicuous 
in our $SNS$ geometry with a single normal plane. Moreover, even in 
the long junction limit ($L=60a \simeq 15 \xi_0$) we find 
$\phi_c \simeq \pi/2$. The non-negligible $\Delta/\mu$ also leads to a 
lowering of $\phi_c$ (and a washing out of the discontinuities in $I(\phi)$ 
at $T=0$), but comparison with non-self consistent calculations, which 
take such normal scattering into account,~\cite{hurd} shows that this 
is negligible compared to the impact of the self-consistency.

The macroscopic phase of the order parameter $\phi(z)$ varies monotonically (i.e., almost 
linearly) across the self-consistently modeled part of the junction. 
However, the plot of $\delta \phi(z)$, obtained after the linear background 
is subtracted from $\phi(z)$, reveals a peculiar spatial distribution which depends 
on the thickness of the junction (Fig.~\ref{fig:phase_sns}). In the short and 
intermediate junction limits, $\delta \phi(z)$ gives a negative contribution 
to $\phi(z)$, which turns into a positive one upon approaching
$I_c$. For thick enough junctions (e.g., $L=20a$ in
Fig.~\ref{fig:phase_sns}) a small bump as the remnant of this 
behavior, persists at the $SN$ boundary, but is completely washed 
out in the long junction limit. Thus, $\delta \phi(z)$  forms a 
``phase anti-dipole'' (i.e., its spatial distribution has positive 
and negative parts opposite to that of the phase dipole, introduced 
in Ref.~\onlinecite{bagwell}), which is a self-consistent response 
to a supercurrent applied in the bulk. From the scaling feature of 
the phase-antidipole we conclude that such counterintuitive behavior 
of the ``phase pile up'' around the $SN$  interface is 
generated by the finite $\Delta/\mu$ effects.

Finally, we examine the local density of states (LDOS) $\rho(\omega,z_i)$ on the
central plane of the $L=10a$ junction, as shown in 
Fig.~\ref{fig:ldos_sns}. At zero Josephson current we find peaks in the LDOS, which are 
of finite width, corresponding to the Andreev bound states~\cite{kulik} (ABS). Moreover, 
instead of a non-zero LDOS all the way to the Fermi energy at $\omega=0$ 
(vanishing linearly as $\omega \rightarrow 0$), which stems from 
quasiparticles traveling almost parallel the $SN$ boundary, a 
minigap $E_g \sim \Delta^2/\mu$ is found  which appears to be the 
consequence of finite $\Delta/\mu$ induced scattering.~\cite{hurd} 
The quantized bound states are the result of an electron (with energy 
below $\Delta$) being retroreflected into a hole at the 
$SN$ interface, while a Cooper pair is injected into the superconductor, 
and vice versa. The hole  is in turn transformed into an electron  
on the opposite surface, so that in the semiclassical picture, a 
bound state forms corresponding to a closed quasiparticle trajectory 
(i.e., an infinite loop of Andreev reflections 
electron$\rightarrow$hole$\rightarrow$electron \ldots). The time-reversed ABS
carries current in the opposite direction, and the two bound states 
are degenerate and decoupled (if there is no interface scattering). When the 
phase gradient is set within the $S$ leads, a phase change appears across the 
junction (i.e., DC Josephson current), and the degenerate ABS split due to the 
Doppler shift. On the other hand, the minigap changes only slightly with 
increasing $\phi$. The two Doppler split peaks 
drift apart monotonically until a bulk phase gradient corresponding to $I_c/2$, when one 
of them reaches the BCS gap edge, while the other one approaches the 
minigap edge. The motion of the ABS for larger current becomes more 
intricate and is described in detail elsewhere.~\cite{review}



\begin{references}
\bibitem{footnote} Mesoscopic superconductivity studies have been given 
a particularly strong impetus through the fabrication of Nb/InAs based junctions, 
where the Fermi level is pinned in the conduction band, thus avoiding a Schottky 
barrier and leading to much higher probability for Andreev reflection.
\end{references}
\end{document}